\documentclass[printer]{aa}
\usepackage{txfonts}
\usepackage{graphicx}

\begin{document}
   \title{Double or binary: on the multiplicity
          of open star clusters} 
   \author{R. de la Fuente Marcos
           \and
           C. de la Fuente Marcos}
   \authorrunning{R. de la Fuente Marcos \and C. de la Fuente Marcos}
   \titlerunning{Optical double or binary open clusters?}
   \offprints{R. de la Fuente Marcos, \email{raul@galaxy.suffolk.es}
              }
   \institute{Suffolk University Madrid Campus, C/ Vi\~na 3,
              E-28003 Madrid, Spain}  
   \date{Received 8 April 2009 / Accepted XX XXXXXXXX XXXX}

   \abstract
      {Observations indicate that the fraction of potential binary
       star clusters in the Magellanic Clouds is about 10\%. In 
       contrast, it is widely accepted that the binary cluster 
       frequency in the Galaxy disk is much lower. 
       }  
      {Here we investigate the multiplicity of clusters in the
       Milky Way disk to either confirm or disprove this dearth 
       of binaries. 
       }
      {We quantify the open cluster multiplicity using complete,
       volume-limited samples from WEBDA and NCOVOCC.
       } 
      {At the Solar Circle, at least 12\% of all open clusters 
       appear to be experiencing some type of interaction with 
       another cluster; i.e., are possible binaries. As in the 
       Magellanic Clouds, the pair separation histogram hints at a 
       bimodal distribution. Nearly 40\% of identified pairs are
       probably primordial. Most of the remaining pairs could 
       be undergoing some type of close encounter, perhaps as a 
       result of orbital resonances. Confirming early theoretical 
       predictions, the characteristic time scale for destruction 
       of bound pairs in the disk is 200 Myr, or one galactic orbit.
       }
      {Our results show that the fraction of possible binary 
       clusters in the Galactic disk is comparable to that in the 
       Magellanic Clouds. 
       }


   \maketitle

   \section{Introduction}
      In a little known paper, Rozhavskii et al. (1976) claimed that
      the fraction of multiple systems among open clusters in the 
      Milky Way was roughly 20\%. This largely neglected result was 
      obtained before any research was undertaken to assess the 
      importance of cluster binarity in the Magellanic Clouds. The 
      firsts of such studies (LMC, Bhatia \& Hatzidimitriou 1988; 
      SMC, Hatzidimitriou \& Bhatia 1990) inferred that the fraction 
      of star clusters in pairs was nearly 10\%. Slightly higher 
      values, 12\%, were found by Pietrzy\'nski \& Udalski (2000). 
      These results were later confirmed by a more rigorous, extensive 
      work completed by Dieball et al. (2002). In contrast, and for 
      the Galaxy, the only widely accepted double or binary cluster 
      system is the $h + \chi$ Persei pair (NGC 869/NGC 884), although 
      the actual physical separation between the pair members is 
      $>$200 pc (see, e.g., de la Fuente Marcos \& de la 
      Fuente Marcos 2009). Nevertheless, Subramaniam et al. (1995) 
      were able to identify a number of additional candidates, 
      concluding that about 8\% of open clusters may be actual 
      binaries, challenging the traditional view. In spite of this
      result, it is still emphasized (see, e.g., Dieball \& 
      Grebel 1998; Bekki et al. 2004) that the number of cluster 
      pairs in the Milky Way is small compared to that in the Clouds. 
      This is usually interpreted as supporting evidence for the higher 
      formation efficiency of bound stellar groups in the Clouds. In 
      all these studies, projected, not three-dimensional, distances 
      were used. 

      In reality, star clusters which have small angular separations 
      are not necessarily physically associated. In most cases, they 
      happen by chance to lie nearly along the same line of sight; 
      they are merely 'optical doubles'. This is to be expected, as 
      clusters are born in complexes (Efremov 1978) which are observed 
      in projection, with some objects formed along shock fronts 
      induced by supernovae. These complexes may also be inclined with 
      respect to the galaxy disk (e.g. the Gould Belt) and this fact
      is customarily neglected in Magellanic Cloud studies. In this 
      context and without information on three-dimensional separations, 
      misidentification of large numbers of double clusters as binaries 
      is, statistically speaking, highly likely. These preliminary 
      considerations pose some obvious questions; if 3-D data are used, 
      is there a real scarcity of binary clusters in our Galaxy with 
      respect to the Magellanic Clouds? How similar are their binary 
      cluster fractions? and their pair separation distributions? Here 
      we attempt to provide statistically robust answers to these and 
      other questions regarding the multiplicity of open clusters. 
      This Letter is organized as follows: in Section 2, we discuss 
      possible formation channels for binary/double clusters. In 
      Section 3, we provide a quantitative analysis of the Galactic 
      binary cluster frequency using samples of open clusters. 
      Individual candidates are considered in Section 4. Results are
      discussed and conclusions summarized in Section 5.

   \section{Double or binary: formation channels}
      Pairs, triplets or higher multiplicity star clusters may form as 
      a result of a number of processes:

      a) Simultaneous formation. Open cluster binaries formed inside a 
         star complex and out of the same molecular cloud are expected 
         to share a common space velocity, but also have very similar 
         ages and chemical composition. They may also be the final 
         products of multiple mergers of smaller clusters. Coformation
         scenarios have been suggested by Fujimoto \& Kumai (1997) and
         Bekki et al. (2004). Objects formed are genetic pairs and 
         therefore true binaries if bound. 

      b) Sequential formation. In this scenario, stellar winds or 
         supernova shocks generated by one cluster induce the collapse 
         of a nearby cloud, triggering the formation of a companion 
         cluster (see, e.g., Brown et al. 1995; Goodwin 1997). 
         A finite but relatively small age difference should be 
         observed. Post-supernova metal contamination may produce 
         different metallicity. If the companion is born out of the 
         same cloudlet forming the originator cluster, common 
         kinematics is expected; if the shocked cloud is unrelated, 
         the kinematics may be different. Pairs formed may be binaries
         or not.

      c) Tidal capture. Open cluster binaries formed by tidal capture 
         must share a common space velocity, but their ages and chemical 
         composition are expected to be rather different. Details of the
         actual mechanism are discussed in, e.g., van den Bergh 
         (1996). Even if they are true binaries, they are not primordial.

      e) Resonant trapping. Dehnen (1998) pointed out that orbital 
         resonances may be responsible for most stellar moving groups 
         present in the solar neighbourhood. This idea was further 
         explored by De Simone et al. (2004), Quillen \& Minchev (2005), 
         Famaey et al. (2005) and Chakrabarty (2007) for the case of 
         resonances induced by the Galactic bar and spiral structure.
         Extrapolating this scenario to entire clusters, we may have 
         formation of double or multiple clusters as a result of 
         resonances of the non-axisymmetric component of the Galactic 
         potential. Cluster pairs must share a common kinematics with 
         rather different ages and metallicities. However, they are not 
         true binaries but pseudo-binaries, transient pairs or multiples. 
         Tidal capture may operate within resonant trapping regions.

      f) Optical doubles: hyperbolic encounters. As in the stellar 
         case, optical doubles are not physical binaries. Although they 
         occupy a limited volume of space (following our criterion, see 
         below, one sphere of radius 15 pc), they do not share common 
         kinematics, but they may have similar ages and metallicities, 
         especially if they are young and belong in the same star complex. 
         Given their small separation, they may well be undergoing some 
         type of close, even disruptive, gravitational encounter.

   \section{Double or binary? the data and the evidence}
      In this section we describe the databases and the classification
      scheme used. The basic criterion used in our preliminary selection is 
      purely dynamical: the pair physical (not projected) separation must 
      be less than three times the average value of the tidal radius for 
      clusters in the Milky Way disk (10 pc, Binney \& Tremaine 2008). 
      Following Innanen et al. (1972), for two clusters separated by a 
      distance larger than three times the outer radius of each cluster, the 
      amount of mutual disruption is negligible. On the other hand,
      general open cluster samples are biased against older clusters because 
      they contain less luminous stars. If any reliable conclusions on the 
      open cluster binary frequency have to be obtained, a complete sample 
      must be used. Completeness of general open cluster samples has been 
      customarily approached assuming uniform surface density in the solar 
      neighborhood (Battinelli \& Capuzzo-Dolcetta 1989, 1991). Piskunov et 
      al. (2006) have concluded that, assuming uniform density, the 
      completeness limit for clusters of any age could be 0.85 kpc. In the 
      following, we consider this group of clusters, forming a volume limited 
      sample, as the best sample.  

         {\it WEBDA.} The {\it Open Cluster Database} \footnote{http://www.univie.ac.at/webda/} 
         (WEBDA, Mermilliod \& Paunzen 2003) is one of the most widely used open
         cluster databases. The latest update of WEBDA (April 2009, Paunzen \& 
         Mermilliod 2009) includes 1756 objects. The number of clusters with both
         age and distance in the database is 1051 (59.9\%). Out of the resulting 
         551,775 pairs, the number of systems with separation $\leq$ 30 pc is 34
         (see Table \ref{binaries1}). Therefore and for the general sample, the 
         fraction of candidate clusters involved in some type of close range 
         dynamical interaction is 6.2\%. If we restrict the analysis to the best 
         sample, 281 objects are included with 39,340 pairs. Out of this best sample, 
         the number of pairs with separation $\leq$ 30 pc is 19 or 12.4\%. Hence, 
         and based only on spatial considerations (as in Magellanic Clouds studies), 
         our best value for the fraction of multiple clusters in the Galactic disk is 
         comparable to that in the Clouds. Most of these cluster pairs have proper 
         motions in WEBDA, with a smaller number also having radial velocities. A very 
         small percentage of open clusters in pairs have known metallicity. 

         {\it NCOVOCC.} The {\it New Catalogue of Optically Visible Open Clusters and Candidates} 
         \footnote{http://www.astro.iag.usp.br/$\sim$wilton/} (NCOVOCC, Dias et al. 
         2002) is also widely used in open cluster studies. The February 2009 version 
         (v2.10, Dias 2009) of NCOVOCC includes 1787 clusters and 982 of these 
         (54.9\%) have known distances and ages, generating 481,671 pairs. After 
         performing the same type of analysis that we did on WEBDA data, we obtain
         the following values for the binary fractions: 5.3\% and 11.4\%. Now we have 
         a total of 27 pairs with 14 in the best sample (see Table \ref{binaries2}).

         The values obtained for the binary fraction of a sample of open clusters in 
         the solar neighbourhood using WEBDA and NCOVOCC are statistically
         consistent and they include the effect of presumed triple systems. Our best 
         value for the open cluster binary fraction in the solar neighbourhood 
         (complete sample) and, therefore in the Galactic disk, is $\sim$12\%. This 
         value matches that of the Magellanic Clouds. If results from both databases 
         are combined, the fraction of possible binary clusters is nearly 15\%. The 
         number of pairs common to WEBDA and NCOVOCC is 18 (ASCC 16 $\equiv$ 
         Brice\~no 1 and ASCC 50 $\equiv$ Alessi 43) with an additional 3 clusters in 
         common but assigned to different companions. Evidence for systems with a
         multiplicity higher than 2 is weak, with a few possible triple or even 
         quadruple systems. The average distances between clusters in the best sample 
         for WEBDA and NCOVOCC are 765.1$\pm$1.8 pc and 754$\pm$2 pc, respectively.
         Our list in Table \ref{binaries1} has only two pairs (\#2 \& \#27) in common
         with Subramaniam et al. (1995); Table \ref{binaries2} shares only 1 pair (\#21). 

\begin{figure}[t]
        \resizebox{\hsize}{!}
         {\includegraphics{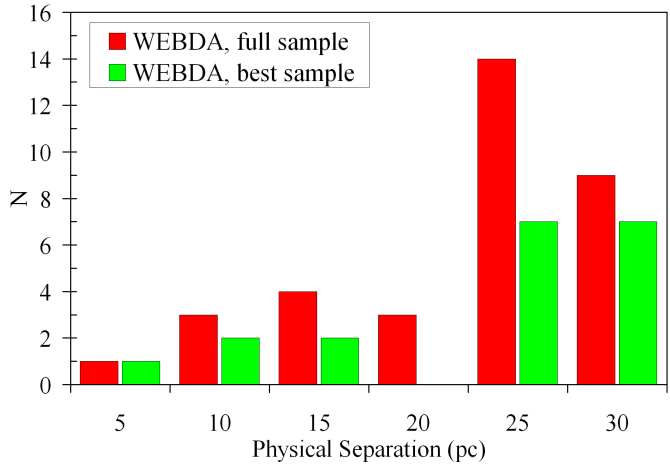}}
        \caption{Pair separation histogram. WEBDA data, Table \ref{binaries1}. 
                }
        \label{bimodal}
     \end{figure}

\begin{figure}[t]
        \resizebox{\hsize}{!}
         {\includegraphics{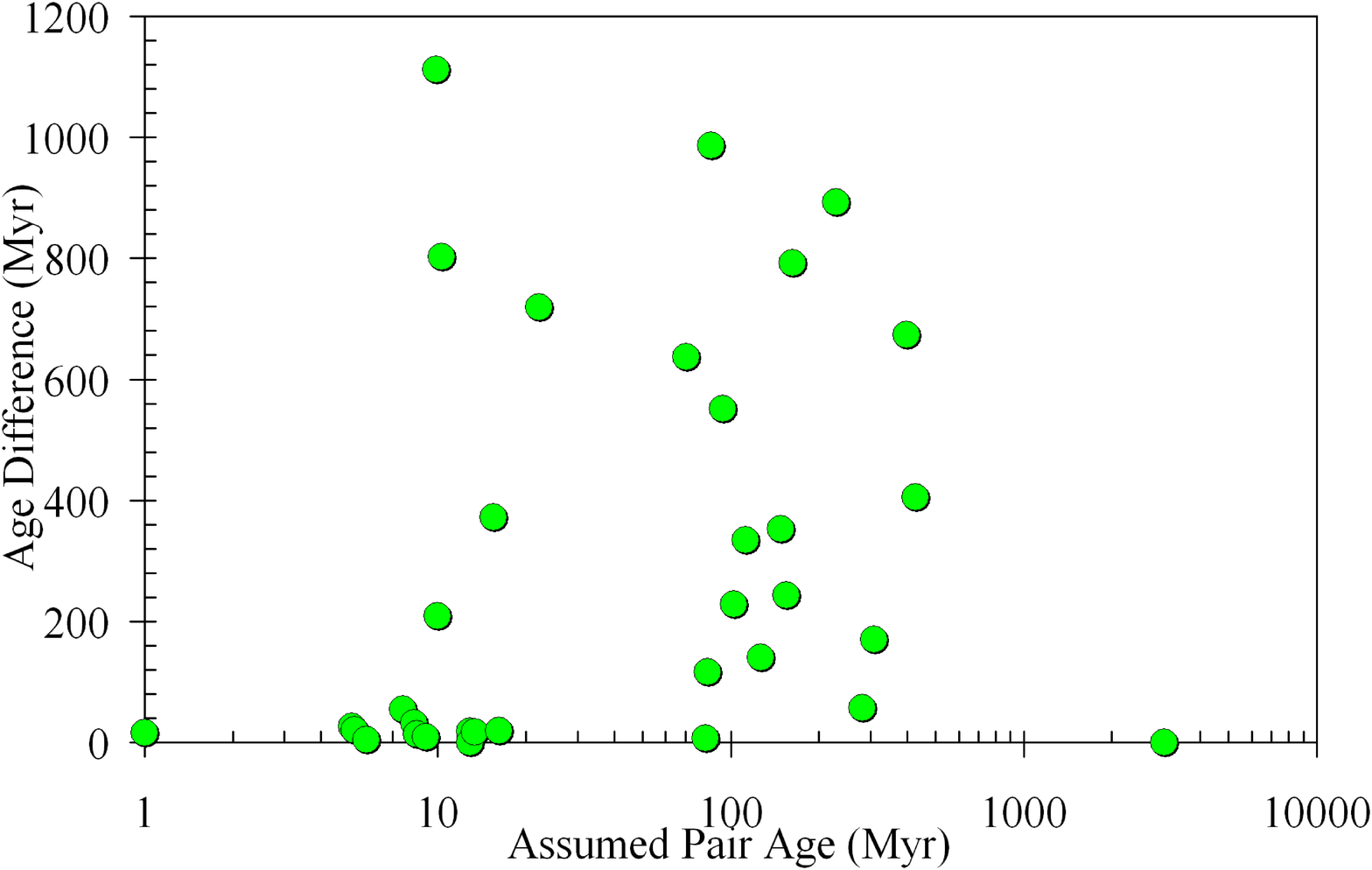}}
        \caption{Age difference of open cluster pairs as a function of their
                 age. WEBDA data. 
                }
        \label{origin1}
     \end{figure}

\begin{figure}[t]
        \resizebox{\hsize}{!}
         {\includegraphics{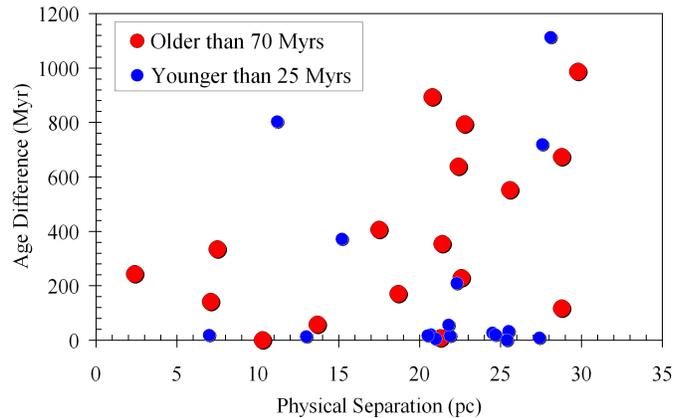}}
        \caption{Age difference of open cluster pairs as a function of their
                 physical separation. WEBDA data. 
                }
        \label{origin2}
     \end{figure}

         The distribution of separations between open clusters in Table \ref{binaries1}
         shows an apparent peak at 10 pc (see Fig. \ref{bimodal}). Similar peaks are
         observed for the LMC ($\sim$6 pc) and the SMC ($\sim$11 pc) but not in
         Subramaniam et al. (1995) for open clusters or in NCOVOCC data. In spite of the
         low number of cluster pairs, the peak seems to be statistically robust as it 
         appears for both the general and best samples. On the other hand, the age of a 
         given cluster pair or multiple system is the age of its youngest member. If we 
         represent the age difference between clusters in a pair as a function of the age 
         of the pair (see Fig. \ref{origin1}), we observe that the vast majority, except three, 
         of almost coeval pairs (age difference $<$80 Myr) are younger than 25 Myr. This 
         implies that primordial pairs do not survive for long and is consistent with 
         similar findings for the LMC and the SMC (Hatzidimitriou \& Bhatia 1990). If the 
         unusual pair \#31 is neglected, no primordial pairs older than 300 Myr exist 
         in the disk, which is also consistent with theoretical expectations discussed in 
         Innanen et al. (1972). Either way, the number of candidate binary clusters among 
         young ones in the best sample is almost 7 times larger than for the general age 
         group. An unexpected age gap is observed for the age range 25-70 Myr. This could 
         be the result of orbit expansion (and subsequent pair destruction) due to mass 
         loss, as observed in $N$-body simulations by Portegies Zwart \& Rusli (2007), but 
         also of early cluster disruption after gas expulsion (Hills 1980). If we further 
         analyze the cluster pairs in these two clearly distinct age ranges ($\leq$25 Myr 
         and $>$70 Myr), we observe some obvious trends (see Fig. \ref{origin2}). The vast 
         majority of primordial pairs appear to have separations in the range 20-30 pc. 
         Most cluster pairs with separations $<$20 pc appear to be old and/or non-primordial 
         and they may have formed by resonant trapping. Notable exceptions are pair \#3 
         (perhaps part of a primordial triple, see below), pair \#7 (a likely long-lived 
         binary) and pair \#31 (another long-lived binary, if real). Older pairs appear to 
         exhibit two disparate groupings: all pairs with separation $<$20 pc have age 
         differences $<$400 Myr or two galactic rotations at the Solar Circle. Wider pairs 
         show random age differences. This may well be a characteristic time scale for 
         disruption of close transient pairs formed by resonant trapping.
%
%-------------------------------------------------------------------------
%
%
%-------------------------------------------------------------------------
%

%
%-------------------------------------------------------------------------
%
%
%-------------------------------------------------------------------------
%

%
%-------------------------------------------------------------------------
%
%
%-------------------------------------------------------------------------
%

   \section{Candidates}
      In Tables \ref{binaries1} and \ref{binaries2} we compile a preliminary list of 
      candidate binary open clusters from WEBDA (34 pairs) and NCOVOCC (27 pairs), 
      respectively. In these tables, pairs are sorted in order of increasing heliocentric
      distance; the horizontal line separates clusters in the complete or best sample. Clusters 
      in both databases appear in boldface. 
       
      i) Binary open clusters.
         In order to classify a pair as binary we restrict the age difference to $\leq$
         50 Myr and having similar kinematics. In WEBDA, clear candidates are pairs
         \#6, \#7, \#18, \#30, and \#34. All of them have similar ages, radial velocity and
         proper motions. Pair \#7 (Loden 1171/Loden 1194) is interesting as the clusters 
         appear to have been able to survive as a bound pair for more than one entire orbit 
         around the Galactic centre. Additional possible binaries with not enough kinematic
         information are pairs \#19, \#27 and \#30. The unstudied pair Carraro 1/Loden 165 is 
         rather peculiar as both objects have intermediate age. Assuming that they are real 
         clusters, it is hard to understand how they have been able to survive as a pair for 
         such a long period of time; they may well be a resonant pair or a close encounter.
         Another obvious primordial binary is the NCOVOCC pair \#24 (Bica 1/Bica 2), which is 
         coeval, has similar proper motion and with a separation of only 3 pc.

      ii) Triple open clusters.
         The system NGC 1981/NGC 1976/Collinder 70 (\#2 and \#3) appears to be a bona fide 
         triple candidate using WEBDA data. These three clusters may constitute a hierarchical 
         triple system with the inner pair being NGC 1981/NGC 1976 with Collinder 70 orbiting 
         around them. NCOVOCC indicates that the triple may be NGC 1976/$\sigma$ Ori/Collinder 70 
         (\#3 and \#4). If we consider both databases simultaneously, the system may well be a 
         quadruple cluster. WEBDA pairs \#8/\#9 and \#10/\#12 cannot be primordial triples 
         but perhaps transient pseudo-triples.

      iii) Tidal capture/resonant trapping.
         Using the available information, it is currently impossible to distinguish pairs 
         formed by tidal capture from those resulting from resonant trapping. On the other
         hand, tidal capture via three-body encounters within resonant trapping regions
         is also a possibility and resonant trapping may be a first step towards tidal
         capture. WEBDA pairs to be included in this group are \#5, \#8, \#9, \#10 or \#12.

      iv) Hyperbolic encounters.
         In this group we include those pairs that are physically close but with disparate kinematics.
         Clear examples in WEBDA are pairs \#4, \#20, \#23 or \#32. Pair \#1 is regarded as
         controversial: the member objects may not be true clusters. The same concern applies
         to many other ASCC candidates. 

   \section{Discussion and conclusions}
      In this Letter, we have shown that binary open clusters appear to constitute
      a statistically significant sample and that the fraction of possible binary 
      clusters in the Galactic disk is comparable to that in the Magellanic Clouds. 
      The spatial proximity of two almost coeval open clusters, compared to the 
      large distances which typically separate these objects, suggests that both 
      objects were formed together. In star-forming complexes, one star cluster might 
      capture another to form a bound state in the presence of a third body or of 
      energy dissipation. This mechanism may also be at work within orbital 
      resonances for non-coeval clusters. However, nearly 40\% of candidate binary 
      clusters that exist in the Milky Way may have been formed in a bound state. Only 
      a small fraction of them, $\sim$17\%, can survive as pairs for more than 25 
      Myr. Our data indicates that binary clusters appear to form both simultaneously 
      and sequentially; multiple clusters form only sequentially. The bimodal nature of the 
      distribution of separations suggests that two formation mechanisms are at work: 
      coformation and resonant trapping. 
    
      How reliable are our present results? Unfortunately, they are affected 
      by the inherent errors associated with the determination of open cluster parameters. 
      The current status of the accuracy of open cluster data has been reviewed by 
      Paunzen \& Netopil (2006). There, they concluded that distances are rather well 
      known because for about 80\% of 395 of the best studied objects, the absolute 
      error is $<$ 20\%. For cluster ages, the situation is the opposite: only 11\% 
      of the best studied objects have errors $<$20\% and 30\% have absolute 
      errors $>$ 50\%. Accordingly, we may assume that our conclusions based
      on distances are probably quite robust but those based on ages may not be as reliable.
      However, they are probably more reliable than those obtained using
      Magellanic Clouds data. A Monte Carlo simulation of 100,000 artificial data sets
      using original WEBDA data altered by errors of 10\%, 20\%, and 30\% in both 
      distance and cluster centre determination gives average binary fractions of 
      10$\pm$2\%, 9$\pm$2\%, and 8$\pm$2\%, respectively, where the error quoted is the
      standard deviation. The binary fraction decreases, as expected, but not dramatically.  
      It is therefore statistically safe to conclude that the open cluster binary fraction 
      is very likely at least 10\%.

   \begin{acknowledgements}
      We would like to thank the anonymous referee for a particularly constructive and 
      quick report. In preparation of this Letter, we made use of the NASA Astrophysics 
      Data System and the ASTRO-PH e-print server. This research has made use of the 
      WEBDA database operated at the Institute of Astronomy of the University of Vienna, 
      Austria. This work also made use of the SIMBAD database, operated at the CDS, 
      Strasbourg, France. 
   \end{acknowledgements}

         \begin{table*}
          \fontsize{8} {10pt}\selectfont
          \tabcolsep 0.10truecm
          \caption{List of candidate binary clusters (WEBDA)}
          \begin{tabular}{rccrrrrrrrrrrr}
           \hline
           \multicolumn{1}{c}{Pair \#}         &
           \multicolumn{1}{c}{Cluster 1}       &
           \multicolumn{1}{c}{Cluster 2}       &
           \multicolumn{1}{c}{$\tau_1$}        &
           \multicolumn{1}{c}{$\tau_2$}        &
           \multicolumn{1}{c}{$\Delta t$}      &
           \multicolumn{1}{c}{$S$}             &
           \multicolumn{1}{c}{$d$}             &
           \multicolumn{1}{c}{$V_{r}$ 1}       &
           \multicolumn{1}{c}{pm RA 1}         &
           \multicolumn{1}{c}{pm dec 1}        &
           \multicolumn{1}{c}{$V_{r}$ 2}       &
           \multicolumn{1}{c}{pm RA 2}         &
           \multicolumn{1}{c}{pm dec 2}        \\
           \multicolumn{1}{c}{}                &
           \multicolumn{1}{c}{}                &
           \multicolumn{1}{c}{}                &
           \multicolumn{1}{c}{(Myr)}           &
           \multicolumn{1}{c}{(Myr)}           &
           \multicolumn{1}{c}{(Myr)}           &
           \multicolumn{1}{c}{(pc)}            &
           \multicolumn{1}{c}{(pc)}            &
           \multicolumn{1}{c}{(km s$^{-1}$)}   &
           \multicolumn{1}{c}{(mas yr$^{-1}$)} &
           \multicolumn{1}{c}{(mas yr$^{-1}$)} &
           \multicolumn{1}{c}{(km s$^{-1}$)}   &
           \multicolumn{1}{c}{(mas yr$^{-1}$)} &
           \multicolumn{1}{c}{(mas yr$^{-1}$)} \\
           \hline
            1 & {\bf ASCC 100} & {\bf ASCC 101} & 102 & 331 & 229 & 23 & 350 
              & -22.89 & 2.29 & -1.51 & - & 0.81 & 1.87 \\ 
            2 & {\bf Collinder 70} & NGC 1981 & 5 & 32 & 26 & 24 & 391
              & 19.49 & 0.36 & -0.68 & 27.94 & 1.01 & 1.14 \\ 
            3 & {\bf NGC 1976} & NGC 1981 & 13 & 32 & 19 & 7 & 400 
              & - & 1.67 & -0.3 & 27.94 & 1.01 & 1.14 \\ 
            4 & {\bf ASCC 20} & ASCC 16 & 22 & 8 & 14 & 13 & 460 
              & 17.62 & -0.09 & 0.51 & 0.75 & 0.75 & -0.18 \\ 
            5 & {\bf NGC 6405} & {\bf ASCC 90} & 94 & 646 & 551 & 26 & 500
              & 10.4 & -2.19 & -5.4 & - & -2.49 & -3.64 \\ 
            6 & {\bf ASCC 21} & {\bf ASCC 18} & 13 & 13 & 0 & 25 & 500
              & 19.9 & 0.52 & -0.62 & 13.39 & 0.89 & -0.02 \\ 
            7 & Loden 1171 & Loden 1194 & 282 & 339 & 57 & 14 & 500
              & - & -13.46 & -5.75 & - & -10.07 & -3.29 \\ 
            8 & {\bf Loden 46} & {\bf ASCC 34} & 1071 & 398 & 673 & 29 & 540 
              & - & -14.11 & 2.98 & - & -2.36 & -1.87 \\ 
            9 & {\bf Loden 46} & {\bf NGC 3228} & 1071 & 85 & 986 & 30 & 544
              & - & -14.11 & 2.98& - & -15.64 & -0.05 \\ 
           10 & NGC 6469 & Ruprecht 139 & 229 & 1122 & 893 & 21 & 550
              & - & 3.04 & -1.2 & - & -1.1 & -2.19 \\ 
           11 & Johansson 1 & Alessi 8 & 199 & 83 & 116 & 29 & 575
              & - & - & - & - & -6.05 & -7.01 \\ 
           12 & Ruprecht 139 & Bochum 14 & 1122 & 10 & 1112 & 28 & 578
              & - & -1.1 & -2.19 & - & - & - \\ 
           13 & Loden 565 & ASCC 68 & 112 & 447 & 334 & 7 & 650
              & 10.1 & -3.14 & -0.96 & - & -4.81 & 3.71 \\ 
           14 & BH 91 & Ruprecht 89 & 162 & 955 & 793 & 23 & 740
              & - & -6.25 & 4.22 & - & -6.82 & 4.36 \\ 
           15 & {\bf ASCC 4} & {\bf NGC 189} & 219 & 10 & 209 & 22 & 752
              & -9.55 & 0.11 & -1.48 & - & - & - \\ 
           16 & NGC 1746 & NGC 1758 & 155 & 398 & 243 & 2 & 760
              & 2.0 & -1.87 & -3.07 & - & -1.44 & -3.5 \\ 
           17 & {\bf Basel 5} & {\bf NGC 6425} & 741 & 22 & 719 & 28 & 778
              & - & - & - & - & 3.11 & -2.79 \\ 
           18 & {\bf Collinder 197} & ASCC 50 & 13 & 30 & 17 & 20 & 838
              & 33.1 & -9.8 & 7.4 & 17.13 & -6.39 & 3.85 \\ 
           19 & {\bf NGC 6250} & {\bf Lynga 14} & 26 & 5 & 21 & 21 & 865
              & - & -0.19 & -3.3 & - & - & - \\ 
           \hline
           20 & Ruprech 91 & ESO 128-16 & 427 & 832 & 405 & 17 & 900
              & 7.3 & -11.1 & 3.11 & - & -2.57 & 6.1 \\ 
           21 & {\bf NGC 2447} & {\bf NGC 2448} & 387 & 15 & 372 & 15 & 1037
              & 21.7 & -4.85 & 4.47 & 15.0 & -3.8 & 4.69 \\ 
           22 & {\bf Ruprecht 172} & {\bf Biurakan 2} & 813 & 10 & 803 & 11 & 1106
              & - & -0.42 & -3.97 & - & -2.52 & -6.53 \\ 
           23 & {\bf NGC 6242} & {\bf Trumpler 24} & 41 & 8 & 32 & 25 & 1138
              & - & 0 & 1.65 & -4 & -0.87 & -1.27 \\ 
           24 & NGC 2302 & NGC 2306 & 70 & 708 & 638 & 22 & 1182
              & - & - & - & - & -0.98 & 1.91 \\ 
           25 & ASCC 6 & Stock 4 & 148 & 501 & 353 & 21 & 1200
              & -20.0 & -1.02 & -1.18 & - & 1.47 & 1.18 \\ 
           26 & {\bf NGC 6613} & {\bf NGC 6618} & 17 & 1 & 16 & 22 & 1296
              & -14.0 & -1.02 & -1.33 & - & 1.79 & -1.96 \\ 
           27 & {\bf Basel 8} & {\bf NGC 2251} & 126 & 267 & 141 & 7 & 1329
              & - & - & - & 24.7 & - & - \\ 
           28 & {\bf Markarian 38} & {\bf Collinder 469} & 8 & 63 & 55 & 22 & 1471
              & -18.0 & 0.07 & -2.2 & - & - & - \\ 
           29 & {\bf Trumpler 22} & NGC 5617 & 89 & 82 & 7 & 21 & 1516
              & - & - & - & - & -2.04 & -2.32 \\ 
           30 & NGC 6871 & Biurakan 1 & 9 & 18 & 9 & 27 & 1574
              & -7.7 & -2.89 & -5.65 & -8.48 & -3.51 & -6.24 \\ 
           31 & {\bf Carraro 1} & {\bf Loden 165} & 3020 & 3020 & 0 & 10 & 1900 
              & - & - & - & - & - & - \\ 
           32 & NGC 659 & NGC 663 & 35 & 16 & 19 & 25 & 1938
              & - & -2.67 & 1.07 & -32.0 & -1.49 & -2.3 \\ 
           33 & Ruprecht 151 & NGC 2428 & 309 & 479 & 170 & 19 & 2100
              & - & -4.3 & 2.1 & - & -2.02 & 1.12 \\ 
           34 & {\bf NGC 3324} & {\bf NGC 3293} & 6 & 10 & 5 & 21 & 2327
              & - & -7.46 & 3.11 & -13.0 & -7.53 & 3.1 \\ 
           \hline
          \end{tabular}
          \begin{list}{}{}
            {\footnotesize
              \item[] $\tau_i$: cluster age in Myr ($i$ = 1, 2). 
              \item[] $\Delta t = \tau_1 - \tau_2$: age difference in Myr. 
              \item[] $S$: cluster pair spatial separation in pc. 
              \item[] $d$: heliocentric distance to the closest member of the pair in pc. 
              \item[] $V_{r} i$: average radial velocity in km s$^{-1}$. 
              \item[] pm RA $i$: average proper motion in RA ($=\mu_{\alpha}\cos\delta$) mas yr$^{-1}$. 
              \item[] pm dec $i$: average proper motion in Dec ($=\mu_{\delta}$) mas yr$^{-1}$. 
            }
          \end{list}
          \label{binaries1}
         \end{table*}

         \begin{table*}
          \fontsize{8} {10pt}\selectfont
          \tabcolsep 0.10truecm
          \caption{List of candidate binary clusters (NCOVOCC)}
          \begin{tabular}{rccrrrrrrrrrrr}
           \hline
           \multicolumn{1}{c}{Pair \#}         &
           \multicolumn{1}{c}{Cluster 1}       &
           \multicolumn{1}{c}{Cluster 2}       &
           \multicolumn{1}{c}{$\tau_1$}        &
           \multicolumn{1}{c}{$\tau_2$}        &
           \multicolumn{1}{c}{$\Delta t$}      &
           \multicolumn{1}{c}{$S$}             &
           \multicolumn{1}{c}{$d$}             &
           \multicolumn{1}{c}{$V_{r}$ 1}       &
           \multicolumn{1}{c}{pm RA 1}         &
           \multicolumn{1}{c}{pm dec 1}        &
           \multicolumn{1}{c}{$V_{r}$ 2}       &
           \multicolumn{1}{c}{pm RA 2}         &
           \multicolumn{1}{c}{pm dec 2}        \\
           \multicolumn{1}{c}{}                &
           \multicolumn{1}{c}{}                &
           \multicolumn{1}{c}{}                &
           \multicolumn{1}{c}{(Myr)}           &
           \multicolumn{1}{c}{(Myr)}           &
           \multicolumn{1}{c}{(Myr)}           &
           \multicolumn{1}{c}{(pc)}            &
           \multicolumn{1}{c}{(pc)}            &
           \multicolumn{1}{c}{(km s$^{-1}$)}   &
           \multicolumn{1}{c}{(mas yr$^{-1}$)} &
           \multicolumn{1}{c}{(mas yr$^{-1}$)} &
           \multicolumn{1}{c}{(km s$^{-1}$)}   &
           \multicolumn{1}{c}{(mas yr$^{-1}$)} &
           \multicolumn{1}{c}{(mas yr$^{-1}$)} \\
           \hline
            1 & Mamajek 1 & Feigelson 1 & 8 & 4 & 4 & 24 & 97
              & 16.1 & -30.0 & 27.8 & 13.0 & -39.5 & -1.0 \\ 
            2 & {\bf ASCC 100} & {\bf ASCC 101} & 102 & 331 & 229 & 23 & 350 
              & -22.89 & 2.29 & -1.51 & -32.0 & 0.81 & 1.87 \\ 
            3 & {\bf Collinder 70} & $\sigma$ Ori & 9 & 13 & 3 & 17 & 387
              & 19.87 & 0.15 & -0.7 & 29.45 & 1.73 & 0.47 \\ 
            4 & $\sigma$ Ori & {\bf NGC 1976} & 13 & 13 & 0.00 & 26 & 399 
              & 29.45 & 1.73 & 0.47 & 28.94 & 1.67 & -0.3 \\ 
            5 & {\bf ASCC 20} & {\bf Brice\~no 1} & 22 & 8 & 14 & 13 & 450 
              & 22.97 & -0.09 & 0.51 & 0 & 0.75 & -0.18 \\ 
            6 & {\bf NGC 6405} & {\bf ASCC 90} & 94 & 646 & 551 & 26 & 487
              & -7.02 & -2.19 & -5.4 & - & -2.49 & -3.64 \\ 
            7 & {\bf ASCC 21} & {\bf ASCC 18} & 13 & 13 & 0 & 25 & 500
              & 19.77 & 0.52 & -0.62 & 24.4 & 0.89 & -0.02 \\ 
            8 & {\bf Loden 46} & {\bf NGC 3228} & 1071 & 85 & 986 & 29 & 540
              & - & -14.11 & 2.98 & -22.39 & -15.64 & -0.05 \\ 
            9 & {\bf Loden 46} & {\bf ASCC 59} & 1071 & 398 & 673 & 30 & 540
              & - & -14.11 & 2.98 & - & -4.99 & 3.65 \\ 
           10 & Johansson 1 & Alessi 8 & 199 & 141 & 58 & 29 & 570
              & - & -3.85 & -3.34 & - & -6.13 & -5.8 \\ 
           11 & {\bf ASCC 4} & {\bf NGC 189} & 219 & 10 & 209 & 22 & 750
              & -9.24 & 0.11 & -1.48 & - & - & - \\ 
           12 & {\bf Basel 5} & {\bf NGC 6425} & 741 & 22 & 719 & 28 & 766
              & - & 3.17 & 1.76 & -3.46 & -3.11 & -2.79 \\ 
           13 & {\bf Collinder 197} & {\bf Alessi 43} & 13 & 30 & 17 & 20 & 838
              & 33.1 & -9.8 & 7.4 & 17.13 & -6.39 & 3.85 \\ 
           14 & {\bf NGC 6250} & {\bf Lynga 14} & 26 & 5 & 21 & 21 & 865
              & -8.04 & -0.19 & -3.3 & - & -3.04 & -3.93 \\ 
           \hline
           15 & {\bf NGC 2447} & {\bf NGC 2448} & 387 & 15 & 372 & 15 & 1037
              & 22.08 & -4.85 & 4.47 & 15.0 & -3.8 & 4.69 \\ 
           16 & ESO 132-14 & NGC 5281 & 794 & 14 & 780 & 27 & 1100
              & - & - & - & -18.52 & -5.29 & -3.45 \\ 
           17 & {\bf Ruprecht 172} & {\bf Biurakan 2} & 813 & 10 & 803 & 11 & 1100
              & - & -1.27 & -3.13 & -22 & -2.31 & -3.78 \\ 
           18 & {\bf NGC 6242} & {\bf Trumpler 24} & 41 & 8 & 32 & 25 & 1131
              & - & 0.38 & -0.19 & -4 & -0.87 & -1.27 \\ 
           19 & NGC 6204 & Hogg 22 & 79 & 6 & 73 & 16 & 1200
              & 53 & 0.04 & -1.45 & -65.2 & -3.16 & -3.77 \\ 
           20 & {\bf NGC 6613} & {\bf NGC 6618} & 17 & 1 & 16 & 22 & 1296
              & -5.4 & -1.02 & -1.33 & -25.3 & 1.79 & -1.96 \\ 
           21 & {\bf Basel 8} & {\bf NGC 2251} & 126 & 267 & 141 & 7 & 1328
              & 11 & 0.43 & -2.13 & 25.33 & -0.24 & -2.6 \\ 
           22 & {\bf Markarian 38} & {\bf Collinder 469} & 8 & 63 & 55 & 22 & 1471
              & -3.2 & -1.76 & -2.12 & - & 0.17 & -1.19 \\ 
           23 & Pismis 19 & {\bf Trumpler 22} & 794 & 89 & 705 & 18 & 1500
              & - & - & - & - & -5.71 & -3.39 \\ 
           24 & Bica 1 & Bica 2 & 4 & 4 & 0 & 3 & 1800
              & - & -0.68 & -1.13 & - & -1.38 & -1.28 \\ 
           25 & {\bf Carraro 1} & {\bf Loden 165} & 3020 & 3020 & 0 & 10 & 1900 
              & - & - & - & - & - & - \\ 
           26 & NGC 2421 & Czernik 31 & 79 & 178 & 98 & 8 & 2200
              & - & -4.26 & 4.89 & - & - & - \\ 
           27 & {\bf NGC 3324} & {\bf NGC 3293} & 6 & 10 & 5 & 21 & 2317
              & -8.5 & -7.46 & 3.11 & -12 & -7.53 & 3.1 \\ 
           \hline
          \end{tabular}
          \label{binaries2}
         \end{table*}

\end{document}